\documentstyle[twoside,fleqn,espcrc2]{article}

\newcommand{\be}{\begin{equation} }
\newcommand{\ee}{\end{equation}  }
\newcommand{\la}{\langle}
\newcommand{\ra}{\rangle}
\newcommand{\beq}{\begin{eqnarray}}
\newcommand{\eeq}{\end{eqnarray}}
\renewcommand{\b}{\beta}

\title {Extension of a new analytical method for
locating  critical temperatures
\thanks{Presented at Lattice'97 (Edinburgh, 22-26 July 1997) }
}

\author { P. Sawicki 
\address{\sl Institute of Physics, Jagiellonian University, \\
ul. Reymonta 4, PL-30-059, Krak\'ow }
\thanks{A fellow of the Foundation for Polish Science}
        } 
\begin{document}

\begin{abstract}
We investigate recently proposed method for locating critical temperatures 
and introduce some modifications which allow to formulate exact criterion 
for any self-dual model. We apply the modified method  for the 
Ashkin-\- -Teller model and show that the exact result for a critical 
temperature is reproduced. We test also a two-layer Ising model 
for the presence of eventual self-duality.
 
\end{abstract}

\maketitle

\section{Introduction}

The recent proposition of the analytical method for locating  
critical temperatures in some spin systems employs
moments of the transfer matrix \cite{wosiek}.
Let us remind briefly its basic assumptions.

For a $d$-dimensional spin system we define the characteristic function
of rank~$n$
\be
\rho_n ( \beta ) = \lim_{L \rightarrow \infty } \left(
\frac{ ( Tr {\cal T} )^n} { Tr {\cal T}^n } \right)
 ^{\frac{1}{L^{d -1} } }
\label{ro}
\ee
where ${\cal T}$ is a transfer matrix of our system with linear size
$L$. The latter quantity was introduced for  regularization purposes.
The dependence on the inverse temperature $\beta=1/T$ is
hidden in the definition of the transfer matrix ${\cal T}$. The same
method relies on the hypothesis  that the location of a maximum of
the function $\rho$ occurs at a critical point (``maximum criterion'') 
\be
\beta_c = \beta_{max}  .  \label{max}
\ee

It remains to test this very attractive hypothesis in practice.
Among others an important question is: how big is the class of models
for which the relation (\ref{max}) indeed holds for any $n$ ?. A conjecture
was given, that a maximum rule is at least valid for self-dual systems.
Indeed it was checked directly  for simple self-dual systems like Ising 
and $q$ state Potts models on a square lattice \cite{sawicki}. 

Recently Souza et al.\ have analyzed the Ashkin-Teller model and found
that the criterion  does not reproduce  correctly its phase structure 
\cite{seixas}. However in particular case of the isotropic Ashkin-Teller,
where the  phase structure is determined from the self-duality
condition, the predicted critical temperature is a good numerical
estimation of the exact value.

In this paper we introduce some modifications which allow to formulate
exact criterion for any self-dual model. We show it on the  Ashkin-Teller
model as an example. Apart from determining critical temperatures our
method can be used to test the existence of duality transformations
for more complicated models. We investigate a certain kind of
duality relation in the two-layer Ising  model  and show that it
is not self-dual.

\section{Application to the Ashkin-Teller model}
The Ashkin-Teller model \cite{baxter} is defined by the Hamiltonian
\beq
H= -J_0 - J \sum_{  \la i,j \ra }   
\sigma_i \sigma_j - J' \sum_{\la i, j \ra } s_i s_j \nonumber \\ 
- J_4 \sum_{ \la i,j \ra} s_i \sigma_i s_j \sigma_j
\eeq
with two Ising-like variables $s_i$ and $\sigma_i$ located on the same
site $i$ of a quadratic lattice. In some sense the AT model
interpolates between the Ising ($J_4=0$) and 4-state Potts model
($J=J'=J_4 $). In general its phase structure is rather complicated 
and can not be simply obtained.

In the simpler isotropic AT model ($J=J'$) the duality transformation
provides for $J \geq J_4 $ the position of a critical point. It
is convenient to introduce dimensionless coefficients
\beq
K_0=\b J_0 ,  & K= \b J , & K_4 =\b J_4
\eeq
and the quantities
\beq
\omega_0 & = & \exp{( 2 K +K_4 +K_0) }, \nonumber \\
\omega_2 & = & \exp{( -K_4 + K_0) },    \nonumber \\
\omega_3 & =& \exp{ (-2 K + K_4 + K_0) }.
\eeq
Then the partition function for the isotropic AT model is invariant
\be
Z(\omega_{0}^{'} , \omega_{2}^{'}, \omega_{3}^{'} ) =
Z(\omega_{0} , \omega_{2}, \omega_{3} ) \label{in}
\ee
under the duality transformation
\beq
\omega_{0}^{'} & = & \frac{1}{2} (\omega_{0} + 2\omega_{2}+\omega_{3}),
\nonumber \\
\omega_{2}^{'} & = & \frac{1}{2} (\omega_{0} - \omega_{3}),
\nonumber \\
\omega_{3}^{'} & = & \frac{1}{2} (\omega_{0} - 2\omega_{2}+\omega_{3}).
\eeq
The critical temperature is known from the self-duality condition
\be
\omega_0 = 2 \omega_2 + \omega_3 . \label{self}
\ee

From  definition (\ref{ro}) follows that
\be
\rho_n (\b)= \lim_{L \rightarrow \infty} \left( \frac{ Z_{1}^{n} }{ Z_n }
\right)^\frac{1}{L}
\ee
where $Z_1$ and $Z_n$ are the partition functions of a chain of length $L$
and of such $n$ coupled chains respectively. In our example the
characteristic function $\rho$ depends on two couplings $K$ and $K_4$ (as can
be easily noticed $K_0$ does not appear in the final result). Therefore
we may expect similar to (\ref{in}) invariance for the function $\rho$
\be
\rho_n (K', K_{4}^{'} ) = \rho_n (K, K_4)
\ee
where $K'$ and $K_{4}^{'}$ are the dual couplings to $K$ and $K_4$
respectively.

For simpler models (like Ising or Potts) the function $\rho$
depends only on one coupling or equivalently on $\b$. Then it can be easily
shown that if there is only one maximum of $\rho$ at some $\b_{max}$ it must
coincide with the self-dual point $\b_{max} =\b^\star$. The same is not true
in our case. In fact the straight line parametrized by
\be
 K= \b J , \; \; K_4= \b J_4    \label{line}
\ee
is not consistent with the duality transformation 
i.e.\ the dual image of a point
$(K, K_4)$ taken from line (\ref{line}) does not lie necessarily on this
line. Our proposition of the modified criterion employs the duality
transformation to find the proper direction for passing the self-dual curve.
Since we are mainly interested in the behavior
of the function $\rho$ near the real critical point it is sufficient
to consider the linearized duality transformation in a vicinity of some
self-dual point $ ( K_{0}^{\star}, K^{\star}, K_{4}^{\star} )$.
Let us denote the vector \linebreak 
$k^T = (K_0 - K_{0}^{\star}, K - K^\star , K_4 -K_{4}^{\star})$ 
and its image under the duality transformation as $k'$. Then
\be
k' = M k
\ee
where the transformation matrix $M$ is:
\be
 \left(
\begin{array}{ccc}

        1  &        \frac{b^3}{a} &       b^2  \\
        0  &       -a^2           &       -ab  \\
        0  &  \frac{a^2 b -b}{a}  &       a^2

\end{array}
\right)
\ee
with $a=\sinh{2 K^\star}$ and $b=\cosh{2 K^\star}$. The matrix $M$ has
three eigenvectors: two with eigenvalue $1$ and one with $-1$. The latter
eigenvector
\be
e_3^T = \left( \frac{2 \cosh{2 K^\star } }{\cosh{4 K^\star} -3 },
\frac{ \sinh{4 K^\star }}{3 -\cosh{4 K^\star} } , 1 \right)
\ee
is interesting since it corresponds to approaching the self-dual curve from
both sides and provides the valid parametrization (analogous to
(\ref{line})) in the modified criterion
\beq
K_4 & = & \frac{3 - \cosh{4 K^\star} }{ \sinh{4K^\star} } (\b - K^\star)
         +K_{4}^{\star}, \nonumber \\
K & = & \b,   \label{slope}
\eeq
where $J=1$ is assumed.

Of course the arguments given above are not a rigorous mathematical proof.
In particular the assumption that for modified criterion the function $\rho$
has only one maximum has to be checked directly for a given model.
For the AT model few lowest moments can be easily computed numerically.
Table.1 contains the critical temperatures obtained from the 
$\rho_2$ function
in the original  and modified method. In accord to \cite{seixas} original
criterion is exact only for $J_4=0$ (two decoupled Ising models) and for
$J_4=1$ (four state Potts model). The modified method gives exact results
for $J_4 \leq 1$ up to rounding errors of order of $10^{-8}$.

\begin{table}
\begin{tabular}{c|c|c|c}
\hline
$J_4$ & $\b_{crit}$  &  $\b_{max}$  &
$ { \hat \b_{max} } $ \\
\hline
 0.00 & 0.44068679&  0.44068679& 0.44068679 \\
 0.10 & 0.41215166&  0.41150224& 0.41215165 \\
 0.20 & 0.38799451&  0.38716551& 0.38799451 \\
 0.30 & 0.36721416&  0.36642983& 0.36721416 \\
 0.40 & 0.34910080&  0.34845723& 0.34910080 \\
 0.50 & 0.33313596&  0.33266131& 0.33313596 \\
 0.60 & 0.31893130&  0.31861775& 0.31893130 \\
 0.70 & 0.30618959&  0.30601070& 0.30618958 \\
 0.80 & 0.29467908&  0.29459934& 0.29467908 \\
 0.90 & 0.28421612&  0.28419628& 0.28421611 \\
 1.00 & 0.27465307&  0.27465307& 0.27465307 \\
\hline
\end{tabular}
\caption{
Comparison of different estimations of the critical temperature
 for the isotropic
Ashkin-Teller with $J=1$: $\b_{crit}$ (exact value)
$\b_{max}$ (original criterion), ${\hat \b_{max}  }$ 
           (modified criterion). }
\end{table}

For $J_4 >1$ the self-dual curve (\ref{self}) is not longer a critical line.
In fact the phase structure is more complicated with two second order
phase transitions. Our criterion predicts the existence of only one phase
transition exactly on the self-dual curve.

\section{Application to the two-layer Ising model}

It is very interesting to apply the modified criterion to other models
(not necessarily self-dual) for which the original maximum rule
does not give exact answers. The two-layer Ising model is an important
example. Its Hamiltonian resembles that for the AT model
\be
H= - J \sum_{  \la i,j \ra }   \sigma_i \sigma_j - J \sum_{\la i, j \ra }
s_i s_j - J_2 \sum_i s_i \sigma_i  .
\ee
The main difference is in the form of the interaction term and in the
interpretation of spin variables $s$ and $\sigma$ which are now
assigned to two distinct layers. Neither the duality transformation
nor the exact critical temperature is known for this model.
One can still attempt to determine the slope of the straight line
in the modified criterion from the condition
\be
\b_{crit} = \b_{max} \; \label{bb} .
\ee
A priori, it is not even certain wheather the relation (\ref{bb})
can be satisfied at all. We found that we were able to estimate
the slope numerically for few lowest moments (Table.2). For $\b_{crit}$
$(J=J_4=1)$ we take the value $0.2760$, the result of Monte Carlo simulations
\cite{burda}. Since it is not exact, with the statistical errors on the
last decimal digit, for comparison the results for two other possible
values of $\b_{crit}$ are also given. The resulting slopes depend on $n$
which is exactly what we expect for not self-dual models.
Since we do not have at our disposal sufficiently precise estimations of
the real critical temperature it is premature to present similar values for
other critical points $(K,K_2)$. However it is rather obvious that
the results would not differ qualitatively. 

\begin{table}
\begin{tabular}{c|c|c|c|c}
\hline
$\b_{crit}$ & n=2 & n=3 & n=4 & n=5 \\
\hline
0.2755 & 0.6399  & 0.5969  & 0.5915  & 0.5929  \\
0.2760 & 0.6233  & 0.5746  & 0.5647  & 0.5625  \\
0.2765 & 0.6067  & 0.5526  & 0.5382  & 0.5324  \\
\hline
\end{tabular}
\caption{ Slopes of the lines (see eq. (\ref{slope})) in the modified
criterion for the two-layer Ising model. }
\end{table}

In conclusion, the method of moments is very closely related to 
self-duality of a given model. For any self-dual model
there is a unique direction in the space of parameters which allows
to construct exact criterion. Conversely if the direction does not 
depend on $n$ then this is a strong argument for the existence of 
a self-duality relation in a model. 

The author thanks the Foundation for Polish Science for a fellowship
and J. Wosiek for useful discussions. This work is supported in part 
by the KBN grant: 2P03B04412.

\end{document}